\documentclass{article}
\usepackage{spconf,cite,graphicx,url,times,amsmath,amssymb,acronym,balance}
\usepackage{color,booktabs,tabularx,balance}
\usepackage{textcomp}
\usepackage[table,xcdraw]{xcolor}
\usepackage{lipsum}

\def\BibTeX{{\rm B\kern-.05em{\sc i\kern-.025em b}\kern-.08em
		T\kern-.1667em\lower.7ex\hbox{E}\kern-.125emX}}

\acrodef{ACD}{Audio content detection}
\acrodef{STFT}{short-time Fourier transform}
\acrodef{SNR}{signal-to-noise ratio}
\acrodef{segSNR}{segmental signal-to-noise ratio}
\acrodef{SRR}{signal-to-reverberation ratio}
\acrodef{PDF}{probability density function}
\acrodef{WER}{word error rate}
\acrodef{SPP}{speech presence probability}
\acrodef{DNN}{deep neural network}
\acrodef{RNN}{recurrent neural network}
\acrodef{CNN}{convolutional neural network}
\acrodef{FC}{fully connected}
\acrodef{CRN}{convolutional recurrent network}
\acrodef{LSTM}{long-term short-term}
\acrodef{GRU}{gated recurrent unit}
\acrodef{FF}{feed forward}
\acrodef{ReLU}{rectified linear unit}
\acrodef{DL}{deep learning}
\acrodef{MAC}{multiply-accumulate}
\acrodef{DNS}{deep noise suppression}
\acrodef{SED}{sound event detection}
\acrodef{ASR}{automatic speech recognition}
\acrodef{CLAP}{Contrastive Audio Language Pretraining}
\acrodef{LLM}{Large Language Model}
\acrodef{VAE}{Variational Auto-Encoder}
\acrodef{GAN}{Generative Adversarial Network}
\acrodef{mrSTFT}{multi-resolution STFT}
\acrodef{FAD}{Fr\'echet Audio Distance}
\acrodef{mAP}{mean Average Precision}
\acrodef{KLD}{Kullback-Leibler Divergence}

\definecolor{darkgreen}{rgb}{0.0, 0.5, 0.0}

\begin{document}
\ninept
\setlength{\abovedisplayskip}{5pt}
\setlength{\belowdisplayskip}{5pt}
\setlength{\abovedisplayshortskip}{3pt}
\setlength{\belowdisplayshortskip}{3pt}
\urlstyle{same}

% \title{SHiCo-AVAE: A General purpose audio encoder with compact semantic latent space and high reconstruction quality}
%attempting to shorten this:
% HOW ABOUT CHICOs OR CHICO-SLS? Continuous Hi Compr Semantic (Latent SPace) AVAE
% \title{SALDi-VAE: Semantic Audio Compression with Language Distillation}
\title{SALAD-VAE: Semantic Audio Compression\\ with Language-Audio Distillation}
\name{Sebastian Braun, Hannes Gamper, Dimitra Emmanouilidou}
\address{Microsoft Research, Redmond, WA, USA}

\maketitle

\begin{abstract}
	% Many modern models operate in a learned latent space, which extracts a compressed and structured representation of the data. In the audio domain, Variational Autoencoders are dominant. We propose a new highly compact continuous VAE, which shows on par high reconstruction fidelity, while using higher latent compression, having a stronger latent representation, and using less computational resources than existing models. This is achieved by augmenting the standard VAE principle with contrastive learning and a CLAP embedding loss. Our resulting latent space outperforms all other audio VAEs in terms of several classification tasks, and on top can directly generate captions in zero-shot manner using the trained CLAP projection layer from the additional loss.
    Modern generative and multimodal models increasingly rely on compact latent  representations that trade and balance semantic richness with high-fidelity reconstruction. We introduce SALAD-VAE, a continuous and highly compact semantic Audio Variational Autoencoder, which operates in the frequency domain and achieves state-of-the-art compression with very low latent frame rate (7.8 Hz) while surfacing semantic structure and producing high audio quality. We enhance the standard VAE semantic losses and augmentation, specifically contrastive learning and CLAP-based embedding distillation, enabling it to generalize across diverse audio domains. 
    With a significantly less computational complex architecture than comparable state-of-the-art VAEs, SALAD-VAE matches their reconstruction quality while it consistently outperforms them on a wide range of classification benchmarks. Furthermore, the proposed additional loss function provides a trained CLAP projection layer, which can be used zero-shot audio captioning and classification matching pretrained CLAP audio-text embeddings.
\end{abstract}


\section{Introduction}

Generative models including latent diffusion models \cite{Liu2023,Evans2025} or multimodal language models \cite{Defossez2024,Fu2025} require or benefit from representing audio in a compact latent domain. 
This representation must satisfy two critical requirements. First, it should compress all relevant information while ideally exposing semantic features that are easily accessible for downstream tasks \cite{Ye2025}. Second, if the task involves reconstructing audio, the representation must enable high-fidelity synthesis that preserves acoustical content such as timbre, timing, and dynamics.
These two requirements are often difficult to unify. As a result, most models tend to specialize: those focused on understanding and reasoning typically lack high-quality audio generation capabilities, while models optimized for generation fidelity often sacrifice interpretability and control.

StableAudio Open \cite{Evans2025} uses a compact convolutional \ac{VAE} to encode time-domain audio into a 64-dimensional latent space at 21~Hz, enabling lightweight generation via latent diffusion.
%, with limited semantic conditioning.
%I wrote "limited" since text-to-audio is itself a level of semantic conditioning
%While efficient and scalable, its low temporal resolution and lack of semantic conditioning can limit controllability.  
Music2Latent \cite{Pasini2024} operates in the frequency domain and uses a Consistency model as decoder, a one-step variant of a diffusion model. The model is optimized for efficient end-to-end training and high-fidelity single-step reconstruction.
Both architectures are transformer-based, which limits support for arbitrary-length inputs and also for streaming capabilities, primarily due to fixed context windows and the quadratic memory scaling of self-attention.
RAVE \cite{Caillon2022} adapts the standard \ac{VAE} architecture originally developed for image modeling \cite{Kingma2014a} into a compact and real-time model, over a two-stage training procedure: representation learning followed by adversarial fine-tuning, enabling high-quality synthesis of 48kHz audio. A controllable latent space allows trade-offs between reconstruction fidelity and compactness. However, it is trained on a limited dataset and may not generalize across diverse audio domains and tasks. XCodec \cite{Ye2025} augments the latent space with a semantic embedding, improving alignment between audio and textual semantics in tasks like speech synthesis and music generation. However, this increases the latent space dimensionality, which may impact efficiency and scalability in downstream generative pipelines.

These recent generative models reflect a growing interest in latent audio representations that balance semantic depth and reconstruction quality, often relying on custom architectures tailored to specific tasks. However, this lack of standardization complicates integration with language models and cross-modal systems. Discrete audio codecs \cite{Zeghidour2021,Defossez2023,Kumar2023} offer a more modular alternative, typically using vector quantization in the bottleneck and trained with a combination of signal reconstruction and adversarial feature-matching losses \cite{Kumar2019}. Their ability to produce discrete token sequences makes them particularly well-suited for integration with language models. 

Although discrete codecs often achieve higher compression, they can suffer from greater information loss. In contrast, continuous audio codecs offer general-purpose representations that, while less modular, tend to preserve fine-grained audio details more effectively. As highlighted in the overview study by Mousavi et al.\ \cite{Mousavi2025}, continuous codecs outperform discrete ones in several tasks due to their superior fidelity and reduced information loss.

% In this work, we aim to bridge this gap by developing a continuous latent space codec that achieves both semantic richness and high reconstruction fidelity, while maintaining practical usability and reasonable architectural complexity. 
% Therefore, to advance the field of continuous generic audio codecs, we propose \textit{HICCSAC}??. We boost the semantic representation by adding a augmentation, a contrastive loss and distillation of a text-audio embedding

In this work, we aim to bridge the gap between semantic richness and audio fidelity by developing a continuous latent-space codec that achieves both, while maintaining practical usability and manageable architectural complexity. Therefore, to advance the field of continuous generic audio codecs, we propose \textit{SALAD-VAE}, a \textbf{S}emantic \textbf{A}udio Compression with \textbf{A}udio-\textbf{L}anguage \textbf{D}istillation VAE\footnote{audio examples: \url{https://sebraun-msr.github.io/SALAD-VAE/}}. Our contributions are as follows:
\begin{itemize}
    \item We propose a continuous frequency-domain audio VAE with compressing audio to a latent vector every 128~ms, resulting in a frame rate of 7.8 Hz. 
    \item We improve generalization to various audio domains by augmenting the training phase with polyphonic data and enforcing random degradations to the VAE input, on the fly. 
    \item We propose a contrastive learning technique for audio VAE by utilizing both a contrastive loss and a joint text-audio embedding distillation loss. This process enhances semantic representation and helps with semantic disentanglement. 
    \item By using an additional projection layer from the distilled VAE embeddings back into the joint text-audio (pretrained) space, we expand the typical audio VAE capabilities to captioning and to zero-shot classification.
\end{itemize}

\section{State of the art Variational Autoencoders}
Given an audio signal $\mathbf{x} \in \mathbb{R}_{1\times T}$ of length $T$, we design an encoder $Enc$ and decoder $Dec$ to obtain a compressed latent representation $\mathbf{Z} \in \mathbb{R}_{D\times M}$ with feature size $D$ and $M$ time frames with $M \ll T$
\begin{eqnarray}
	\mathbf{Z} = Enc\{\mathbf{x}\} \\
	\mathbf{\hat{x}} = Dec\{\mathbf{Z}\}
\end{eqnarray}
where $\mathbf{\hat{x}}$ is a reconstructed version of $\mathbf{x}$. The encoder-decoder pair is parameterized by a set of learnable parameters $\theta$.

A \ac{VAE} is trained using a reconstruction loss on the data $\mathbf{x}$ and the \ac{KLD} on the latent space $\mathbf{Z}$. To improve reconstruction quality, it is common to add adversarial and feature matching loss terms. Without these additional losses, the model tends to produce low-pass filtered outputs. The total reconstruction loss is computed as a weighted sum of multi-resolution \ac{STFT} loss, adversarial loss and feature matching loss:
\begin{eqnarray}
	\mathcal{L}_\text{rec}(\mathbf{x}) \!\!\! &=&\!\!\! \mathcal{L}_\text{mrSTFT}(\mathbf{x}, \mathbf{\hat{x}}) + \lambda_\text{adv} \mathcal{L}_\text{adv}(\mathbf{\hat{x}}) +  \lambda_\text{fm} \mathcal{L}_\text{fm}(\mathbf{x}, \mathbf{\hat{x}}) \label{eq:recon_loss}\\
	%	\mathcal{L}_\text{latent} = \mathcal{L}_\text{KL}(\mathbf{z})) \\
	\mathcal{L}_\text{VAE} \!\!\! &=&\!\!\! \mathcal{L}_\text{rec} + \lambda_\text{KL} \mathcal{L}_\text{KL}(\mathbf{Z})  \label{eq:vae_loss}
\end{eqnarray}
where the $\lambda$ factors are scalar weights to balance the losses.
This is state-of-the-art as proposed by several works \cite{Evans2025,Defossez2023,Wu2023,Caillon2022}, where typically a L1 \ac{mrSTFT} loss, a least-squares \ac{GAN} loss is used and a L1 feature matching (fm) objective between all intermediate discriminator features of signal targets and reconstructions.

\section{Proposed methods}
We propose several additional training techniques including data augmentation and losses to the standard \ac{VAE} described above. 

\subsection{Polyphonic augmentation and denoising autoencoder}
\label{sec:data_augmentation}
To improve the generalization, we generate polyphonic data on the fly, in the fashion of \emph{mix-up} \cite{Zhang2018}, by mixing up to $N$ audio sources files. Further, we employ the principle of the \emph{denoising autoencoder} \cite{Bengio2013} by adding random degradations to the VAE input, but not to the training target. The model is therefore encouraged to remove degradations such as bandwidth limitation, codec artifacts and level variations. The augmented input signal is given by
\begin{eqnarray}
	\mathbf{x} = \sum_{n=1}^N \mathcal{A}\{\mathbf{s}_n\}
\end{eqnarray}
where we mix $N$ audio clips $\mathbf{s}_n$, each augmented with a different instance of source augmentation (e.\,g., EQ, reverb, loudness, level jump, time shift, pitch shift).
By applying random microphone signal degradation functions $\mathcal{M}$ to the input, we train the auto-encoder with
\begin{eqnarray}
    \label{eq:denoising_vae}
	\mathbf{\hat{x}} = Dec\left\{ Enc\{\mathbf{y}\} \right\}
\end{eqnarray}
where $\mathbf{y} = \mathcal{M}\{\mathbf{x}\}$, and the \ac{VAE} reconstruction loss is still computed as in \eqref{eq:recon_loss} between $\mathbf{x}$ and $\mathbf{\hat{x}}$.

\subsection{Contrastive semantic loss}
We propose a contrastive learning technique for audio VAE to aid \emph{semantic disentanglement}. For each audio sample, we create two \emph{differently augmented} versions containing the \emph{same content}. Specifically, the same set of source signals $\mathbf{s}_n,\; n \!\in\! \mathcal{S}_\text{pos}$ is used to create two different training input signals $\mathbf{y}_i, \mathbf{y}_j$ by applying different instances of source and mic augmentations $\mathcal{A}, \mathcal{M}$. All other signals in the training batch are considered negative examples. We then use a contrastive loss attracting the latent variables of the two positive augmented versions with same source content $\mathbf{Z}_i,\mathbf{Z}_j$, while repelling the latents of all other signals.
The contrastive loss is given by
\begin{eqnarray}
	\mathcal{L}_\text{contr} = \frac{1}{|\mathcal{B}|} \sum_{i,j \in \mathcal{B}} \log \frac{\exp(\text{sim}(P_c(\mathbf{Z}_i),\, P_c(\mathbf{Z}_j)))}{\sum_{k, k\neq i} \exp(\text{sim}(P_c(\mathbf{Z}_i),\, P_c(\mathbf{Z}_k)))}
\end{eqnarray}
where $P_c$ is a learnable time aggregation and projection module and $\text{sim}(\cdot)$ denotes the cosine similarity. Embeddings are time averaged as we want to contrast only on the time invariant semantic representation level, not on the fine-grained signal level. Projection to a larger space before the contrastive loss has been shown beneficial in \cite{Chen2020,Chen2021}.

\subsection{CLAP loss}
To enhance the semantic representation of the VAE embeddings further, we align an up-projected version with a pre-trained text-audio embedding (here specifically \ac{CLAP} \cite{Elizalde2024}) space by adding a similarity loss:

\begin{eqnarray}
	\label{eq:clap_loss}
	\mathcal{L}_\text{CLAP} = \frac{1}{|\mathcal{B}|} \sum_{i \in \mathcal{B}} 2 - \text{sim}(CLAP(\mathbf{x}_i),\; P_\text{L}(\mathbf{Z}_i))
\end{eqnarray}
where $P_\text{L}(\cdot)$ is a temporal average and projection layer that converts the lower dimensional, time-variant VAE embedding space into the higher dimensional time-invariant CLAP embedding space (1024), such that $\mathbf{z}_\text{CLAP} = CLAP(\mathbf{x}), \; P_\text{L}(\mathbf{Z}) \in \mathbb{R}_{CLAP}$. This essentially distills the text-audio alignment knowledge of CLAP into our embedding space, without requiring paired audio-text data.

\subsection{Overall combined training scheme}
\begin{figure}[tbp]
	\centerline{\includegraphics[width=\columnwidth,clip,trim=65 30 120 30]{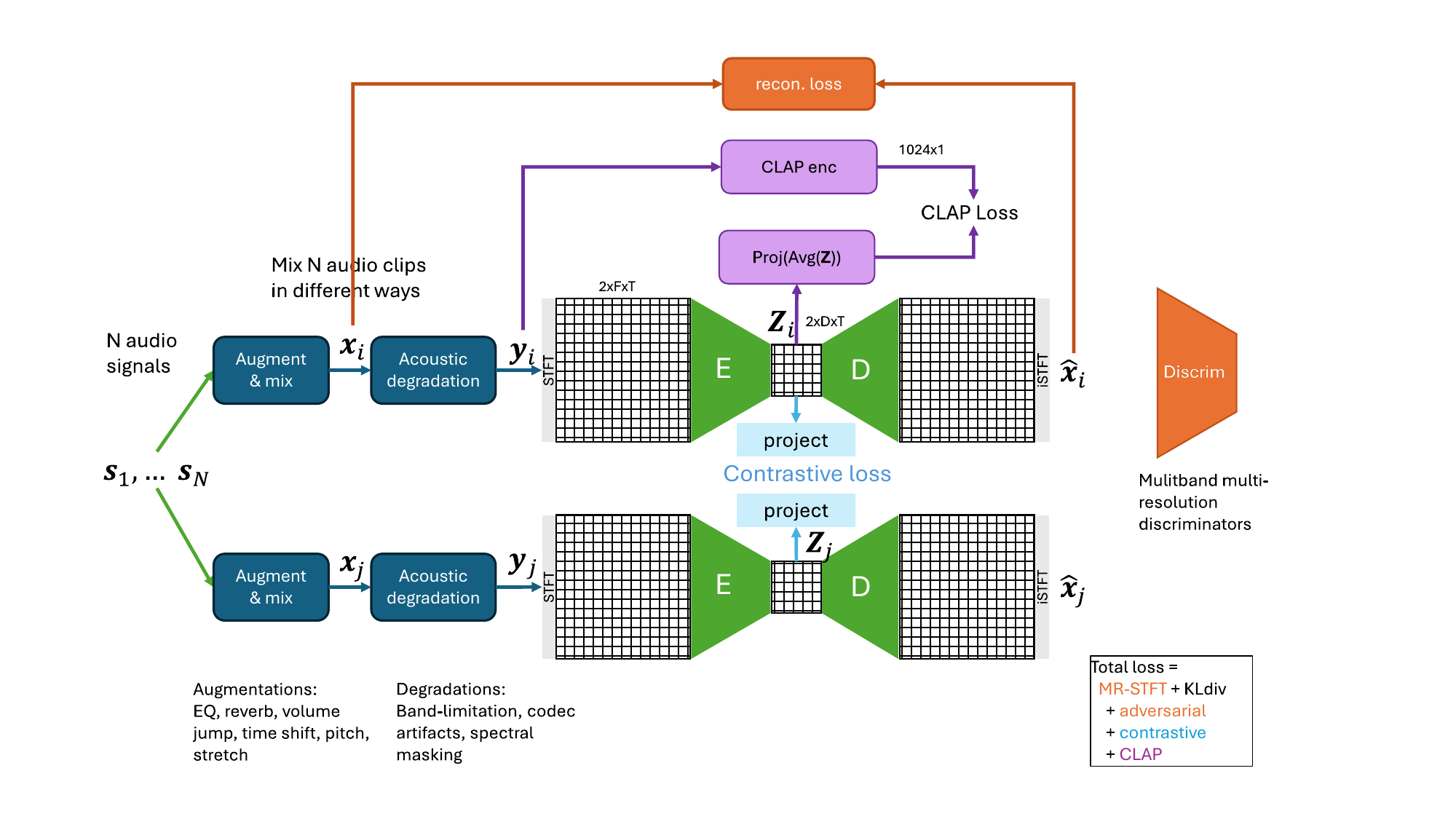}}
	\caption{Proposed training scheme depicting signal augmentations, CLAP and contrastive losses.}
	\label{fig:blockdiag}
\end{figure}
The overall training scheme is depicted in Fig.~\ref{fig:blockdiag}, which is optimized on the overall loss
\begin{eqnarray}
	\mathcal{L}_\text{prop} = \mathcal{L}_\text{rec} + \lambda_\text{KL} \mathcal{L}_\text{KL}(\mathbf{z})) + \lambda_\text{contr} \mathcal{L}_\text{contr} + \lambda_\text{CLAP} \mathcal{L}_\text{CLAP} \label{eq:total_loss}
\end{eqnarray}
We adjust $\lambda_\text{KL}$ with cyclical cosine annealing as proposed in \cite{Fu2019}.

\subsection{Captioning and zero-shot classification}
By distilling \ac{CLAP} language-audio knowledge into the VAE latent space, we enable caption generation directly from our VAE latent representations. We repurpose the projection layer $P_\text{L}$ from \eqref{eq:clap_loss} to map time-averaged VAE embeddings into the CLAP space, which are then decoded using the pre-trained CLAP text decoder (GPT-2).
Further, this alignment also enables zero-shot audio classification capabilities, by selecting the text label with the highest cosine similarity to the projected audio embedding, following~\cite{Elizalde2024}. This allows our VAE to generalize to unseen classes without additional training.

\section{Implementation details}
\subsection{Neural architecture}
\begin{figure}[tbp]
	\centerline{\includegraphics[width=0.5\textwidth,clip,trim=60 80 60 50]{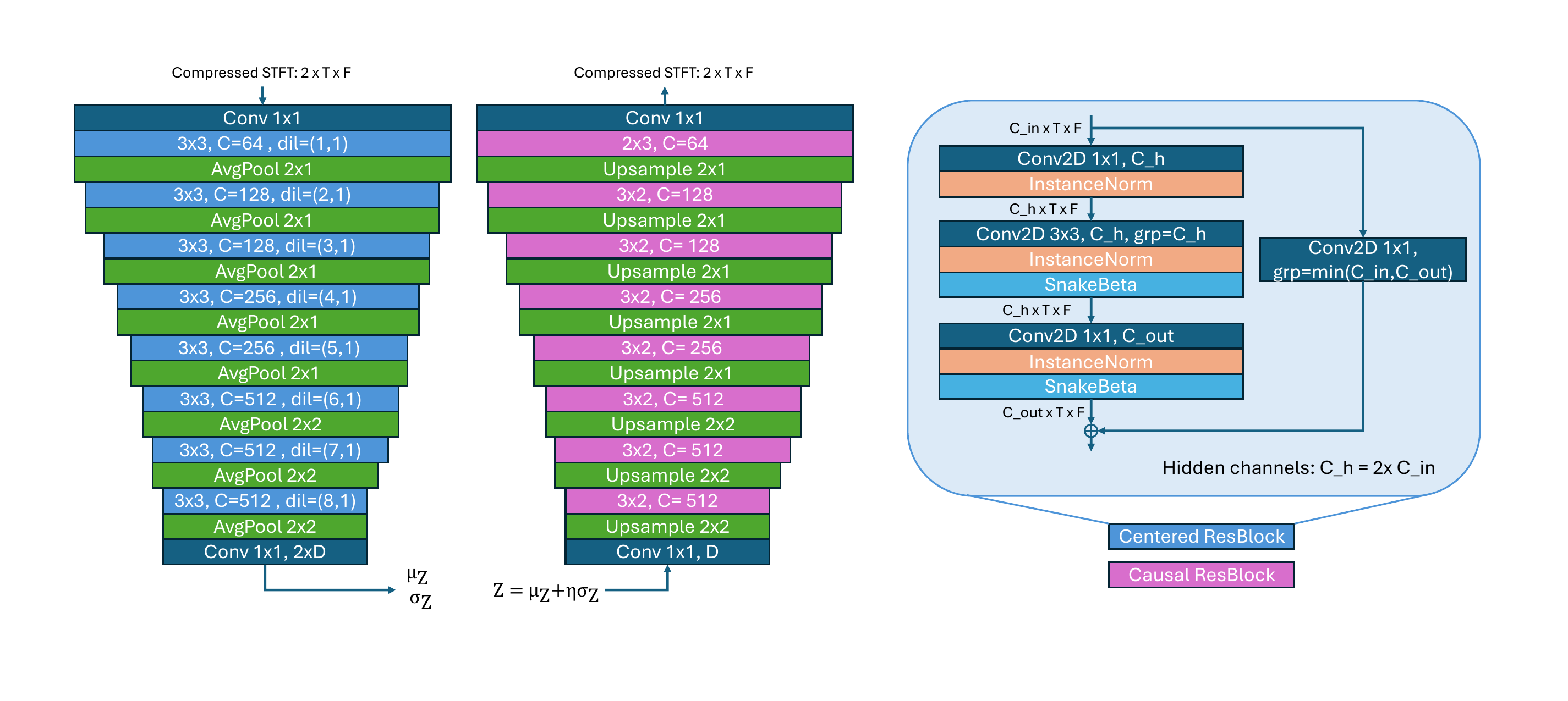}}
	\caption{Left: proposed VAE architecture using centered ResBlocks in encoder and causal ResBlocks in decoder. Right: Inverted bottleneck ResBlock.}
	\label{fig:vae_arch}
\end{figure}

The \ac{VAE} is a fully convolutional model operating in the frequency domain as shown in Fig.~\ref{fig:vae_arch}. 
Input and output features is the power-law compressed \ac{STFT} with frequency and time dimension $F$ and $T$, where the real and imaginary part are treated as convolutional channels. The encoder consists of 8 convolutional blocks with channels [64,128,128,256,256,512,512,512], or [64,128,256,512,512,1024,1024,2048] for the large model. Each conv block is a inverted bottleneck residual layer \cite{Sandler2018}, which projects the features to 2x the size, does a depth-wise convolution and projects it back, wrapped with a 1x1 conv skip connection. While each layer downsamples the frequency dimension, only the inner 3 layers downsample time. The bottleneck is a simple 1x1 conv layer. The decoder is a symmetrically mirrored version of the encoder with respective upsampling layers, using nearest neighbor interpolation to mitigate artifacts. We use instance normalization and SnakeBeta activations \cite{Ziyin2020}, which can improve audio quality due to their symmetricity. We deliberately design the encoder stronger than the decoder to enforce stronger representation learning: the encoder uses centered convolutions and increasing dilation for larger receptive field, while the decoder uses uses only causal convolutions and shorter time kernels.
The receptive field of the encoder is 5.4~s. With the STFT operating on 32~ms windows with 16~ms hop, resulting in a latent frame rate of 128~ms (7.8~Hz).

We use multi-band multi-resolution discriminators. The discriminator architecture follows \cite{Defossez2023}, a 6 layer 2D CNN with 32 channels, kernel size (3,3), stride (2,2). We feed the real and imaginary part of the compressed \ac{STFT} as input, in multiple resolutions, window sizes $[1024, 256, 128]$ and 50\% overlap. We use one set of full-band discriminators and band-split fractions $[0, 0.1, 0.25, 0.5, 0.75, 1]$ of fullband \cite{Jang2021}.

We use a complex \ac{mrSTFT} loss with prime Hann window lengths of [2039, 1021, 503, 257, 127, 61, 31] to better catch periodic artifacts. All STFTs use 75\% overlap and magnitude compression of 0.3 and a L1-norm loss.

The \ac{VAE} is first pre-trained on \ac{mrSTFT} loss and \ac{KLD} (faded in with annealing) to learn to produce some audio. After several epochs, other training losses such as discriminators, CLAP and contrastive loss are added. We train with a batch size of 64, learning rate of 0.001, AdamW optimizer \cite{Loshchilov2019} with betas (0.5,0.99), and exponential moving average (EMA) model weight update \cite{Karras2024} with momentum of 0.9999.

\subsection{Training data}
We train on AudioSet \cite{Gemmeke2017} (5500~h), which contains a large variety of speech, music and sounds. The data is augmented as described in Sec.~\ref{sec:data_augmentation} by randomly cropping and concatenating sequences to obtain 10~s sequences, mixing up to 2 such audio sequences, and applying random EQ, reverb, loudness, level jump, time shift or pitch shift to each audio file as function $\mathcal{A}$. To the mixed audio, we apply random degradations like spectral masking, audio codecs, bandpass filtering, non-linear distortions and level variations as function $\mathcal{M}$.

\section{Experimental results}

% \begin{figure}[tbp]
% 	\centerline{\includegraphics[width=\columnwidth]{loss_ablation.png}}
% 	\caption{Ablation of loss contributions.}
% 	\label{fig:loss_ablation}
% \end{figure}
\begin{table*}[tb]
    \centering
    \vspace{-6pt}
    \caption{Ablation of loss contributions for the proposed VAE.}
    \label{tab:loss_ablation}
% \usepackage{booktabs}
% \usepackage{graphicx}
\resizebox{\textwidth}{!}{%
\begin{tabular}{@{}l|ccc|ccccc|cccc|cc@{}}
\toprule
 &
  \multicolumn{3}{c|}{\cellcolor[HTML]{CAEDFB}reconstruction quality} &
  \multicolumn{5}{c|}{\cellcolor[HTML]{F2CEEF}latent space probing} &
  \multicolumn{4}{c|}{\cellcolor[HTML]{DAF2D0}Zero-Shot classification} &
  \multicolumn{2}{c}{\cellcolor[HTML]{FBE2D5}captioning (SPIDEr)} \\ %\midrule
{\color[HTML]{000000} \textbf{loss}} &
  {\color[HTML]{000000} \textbf{DistillMOS}} &
  {\color[HTML]{000000} \textbf{WER}} &
  {\color[HTML]{000000} \textbf{FAD}} &
  {\color[HTML]{000000} \textbf{Scenes}} &
  {\color[HTML]{000000} \textbf{Events}} &
  {\color[HTML]{000000} \textbf{Emotion}} &
  {\color[HTML]{000000} \textbf{Music}} &
  {\color[HTML]{000000} \textbf{Instrument}} &
  {\color[HTML]{000000} \textbf{Scenes}} &
  {\color[HTML]{000000} \textbf{Events}} &
  {\color[HTML]{000000} \textbf{Music}} &
  {\color[HTML]{000000} \textbf{Instrument}} &
  {\color[HTML]{000000} \textbf{Clotho}} &
  {\color[HTML]{000000} \textbf{AudioCaps}} \\\midrule
\rowcolor[HTML]{63BE7B} 
\cellcolor[HTML]{D9D9D9}CLAP &
  \cellcolor[HTML]{D9D9D9}N/A &
  \cellcolor[HTML]{D9D9D9}N/A &
  \cellcolor[HTML]{D9D9D9}N/A &
  0.54 &
  0.46 &
  0.43 &
  0.83 &
  0.63 &
  0.45 &
  0.53 &
  0.72 &
  0.74 &
  0.27 &
  0.46 \\
chance &
  N/A &
  N/A &
  N/A &
  \cellcolor[HTML]{F8696B}0.10 &
  \cellcolor[HTML]{F8696B}0.01 &
  \cellcolor[HTML]{F8696B}0.25 &
  \cellcolor[HTML]{F8696B}0.10 &
  \cellcolor[HTML]{F8696B}0.10 &
  \cellcolor[HTML]{F8696B}0.10 &
  \cellcolor[HTML]{F8696B}0.01 &
  \cellcolor[HTML]{F8696B}0.10 &
  \cellcolor[HTML]{F8696B}0.10 &
  \cellcolor[HTML]{F8696B}0.00 &
  \cellcolor[HTML]{F8696B}0.00 \\\midrule
\cellcolor[HTML]{D9D9D9}recon+KLD &
  \cellcolor[HTML]{FFEB84}1.26 &
  \cellcolor[HTML]{FDBE7C}0.93 &
  \cellcolor[HTML]{FFEB84}1191 &
  \cellcolor[HTML]{FDD27F}0.29 &
  \cellcolor[HTML]{FDCB7D}0.06 &
  \cellcolor[HTML]{FCBC7B}0.29 &
  \cellcolor[HTML]{FCC57C}0.42 &
  \cellcolor[HTML]{FEE081}0.25 &
  \cellcolor[HTML]{D9D9D9}N/A &
  \cellcolor[HTML]{D9D9D9}N/A &
  \cellcolor[HTML]{D9D9D9}N/A &
  \cellcolor[HTML]{D9D9D9}N/A &
  \cellcolor[HTML]{D9D9D9}N/A &
  \cellcolor[HTML]{D9D9D9}N/A \\
recon+KLD+contrastive &
  \cellcolor[HTML]{F8696B}1.16 &
  \cellcolor[HTML]{F8696B}1.08 &
  \cellcolor[HTML]{FCAF79}1320 &
  \cellcolor[HTML]{FEDD81}0.31 &
  \cellcolor[HTML]{FDD47F}0.07 &
  \cellcolor[HTML]{FFEB84}0.31 &
  \cellcolor[HTML]{FDD07E}0.46 &
  \cellcolor[HTML]{FFEB84}0.27 &
  N/A &
  N/A &
  N/A &
  N/A &
  N/A &
  N/A \\
\cellcolor[HTML]{D9D9D9}recon+KLD+CLAP &
  \cellcolor[HTML]{FCB77A}1.22 &
  \cellcolor[HTML]{FFEB84}0.85 &
  \cellcolor[HTML]{FFDA81}1229 &
  \cellcolor[HTML]{7EC67D}0.51 &
  \cellcolor[HTML]{B0D580}0.27 &
  \cellcolor[HTML]{A0D07F}0.38 &
  \cellcolor[HTML]{7BC57D}0.78 &
  \cellcolor[HTML]{CADC81}0.39 &
  \cellcolor[HTML]{D9D9D9}N/A &
  \cellcolor[HTML]{D9D9D9}N/A &
  \cellcolor[HTML]{D9D9D9}N/A &
  \cellcolor[HTML]{D9D9D9}N/A &
  \cellcolor[HTML]{D9D9D9}N/A &
  \cellcolor[HTML]{D9D9D9}N/A \\
recon+KLD+CLAP+contr &
  \cellcolor[HTML]{F97E6F}1.18 &
  \cellcolor[HTML]{F9726D}1.06 &
  \cellcolor[HTML]{F8696B}1467 &
  \cellcolor[HTML]{75C47D}0.52 &
  \cellcolor[HTML]{C3DA81}0.23 &
  \cellcolor[HTML]{A9D380}0.38 &
  \cellcolor[HTML]{9FD07F}0.72 &
  \cellcolor[HTML]{C1D981}0.41 &
  \cellcolor[HTML]{D8E082}0.30 &
  \cellcolor[HTML]{D8E082}0.29 &
  \cellcolor[HTML]{B9D780}0.63 &
  \cellcolor[HTML]{EBE683}0.33 &
  \cellcolor[HTML]{F6E984}0.10 &
  \cellcolor[HTML]{E4E483}0.22 \\
\cellcolor[HTML]{D9D9D9}recon+KLD+mbGAN &
  \cellcolor[HTML]{63BE7B}2.76 &
  \cellcolor[HTML]{63BE7B}0.17 &
  \cellcolor[HTML]{79C47C}582 &
  \cellcolor[HTML]{FFEB84}0.33 &
  \cellcolor[HTML]{FFEB84}0.08 &
  \cellcolor[HTML]{FDCB7D}0.29 &
  \cellcolor[HTML]{FFEB84}0.55 &
  \cellcolor[HTML]{FEE883}0.26 &
  \cellcolor[HTML]{D9D9D9}N/A &
  \cellcolor[HTML]{D9D9D9}N/A &
  \cellcolor[HTML]{D9D9D9}N/A &
  \cellcolor[HTML]{D9D9D9}N/A &
  \cellcolor[HTML]{D9D9D9}N/A &
  \cellcolor[HTML]{D9D9D9}N/A \\
recon+KLD+mbGAN, no enhance &
  \cellcolor[HTML]{A4D17F}2.14 &
  \cellcolor[HTML]{B0D47F}0.51 &
  \cellcolor[HTML]{C2D980}914 &
  \cellcolor[HTML]{FEDB80}0.30 &
  \cellcolor[HTML]{FDD07E}0.07 &
  \cellcolor[HTML]{FDCB7D}0.29 &
  \cellcolor[HTML]{FDD47F}0.47 &
  \cellcolor[HTML]{FDD27F}0.23 &
  N/A &
  N/A &
  N/A &
  N/A &
  N/A &
  N/A \\
\cellcolor[HTML]{D9D9D9}recon+KLD+CLAP+contr+mbGAN &
  \cellcolor[HTML]{7AC57D}2.55 &
  \cellcolor[HTML]{71C27B}0.23 &
  \cellcolor[HTML]{63BE7B}480 &
  \cellcolor[HTML]{A3D17F}0.46 &
  \cellcolor[HTML]{C7DB81}0.22 &
  \cellcolor[HTML]{D3DF82}0.34 &
  \cellcolor[HTML]{78C47D}0.79 &
  \cellcolor[HTML]{E3E383}0.33 &
  \cellcolor[HTML]{FCBB7A}0.19 &
  \cellcolor[HTML]{FCB679}0.12 &
  \cellcolor[HTML]{FDD880}0.50 &
  \cellcolor[HTML]{FCBA7A}0.20 &
  \cellcolor[HTML]{FEDA80}0.08 &
  \cellcolor[HTML]{FCC47C}0.12 \\ \bottomrule
\end{tabular}%
}
\end{table*}
%
%
% \begin{figure*}[tb]
% 	\centerline{\includegraphics[width=\textwidth]{results_overall.png}}
% 	\caption{Results of proposed system compared to baselines.}
% 	\label{fig:results_overall}
% \end{figure*}
\begin{table*}[t]
    \centering
    \vspace{-15pt}
    \caption{Results of proposed system compared to baselines.}\label{tab:results_overall}
% \usepackage{booktabs}
% \usepackage{graphicx}
\resizebox{\textwidth}{!}{%
\begin{tabular}{@{}ll|ccc|ccccc|cc|ccc@{}}
\toprule
 &
   &
  \multicolumn{3}{c|}{\cellcolor[HTML]{CAEDFB}reconstruction quality} &
  \multicolumn{5}{c|}{\cellcolor[HTML]{F2CEEF}latent space} &
  \multicolumn{2}{c|}{\cellcolor[HTML]{FBE2D5}captioning (SPIDEr)} &
  \multicolumn{3}{c}{\cellcolor[HTML]{DAF2D0}architecture properties} \\ %\midrule
{\color[HTML]{000000} \textbf{model}} &
  {\color[HTML]{000000} \textbf{loss}} &
  {\color[HTML]{000000} \textbf{DistillMOS}} &
  {\color[HTML]{000000} \textbf{WER}} &
  {\color[HTML]{000000} \textbf{FAD}} &
  {\color[HTML]{000000} \textbf{Scenes}} &
  {\color[HTML]{000000} \textbf{Events}} &
  {\color[HTML]{000000} \textbf{Emotion}} &
  {\color[HTML]{000000} \textbf{Music}} &
  {\color[HTML]{000000} \textbf{Instrument}} &
  {\color[HTML]{000000} \textbf{Clotho}} &
  {\color[HTML]{000000} \textbf{AudioCaps}} &
  {\color[HTML]{000000} \textbf{params (M)}} &
  {\color[HTML]{000000} \textbf{GMAC/s}} &
  {\color[HTML]{000000} \textbf{rate (Hz)}} \\\midrule
\rowcolor[HTML]{D9D9D9} 
original audio &
   &
  \cellcolor[HTML]{63BE7B}4.13 &
  \cellcolor[HTML]{63BE7B}0.03 &
  \cellcolor[HTML]{63BE7B}0 &
  N/A &
  N/A &
  N/A &
  N/A &
  N/A &
  N/A &
  N/A &
  N/A &
  N/A &
  N/A \\
CLAP (audio enc only) &
   &
  N/A &
  N/A &
  N/A &
  \cellcolor[HTML]{63BE7B}0.54 &
  \cellcolor[HTML]{63BE7B}0.46 &
  \cellcolor[HTML]{63BE7B}0.43 &
  \cellcolor[HTML]{63BE7B}0.83 &
  \cellcolor[HTML]{63BE7B}0.63 &
  \cellcolor[HTML]{63BE7B}0.27 &
  \cellcolor[HTML]{63BE7B}0.46 &
  \cellcolor[HTML]{D3DE81}32.8 &
  \cellcolor[HTML]{97CD7E}6.8 &
   \\\midrule
\cellcolor[HTML]{D9D9D9}StableAudio Open VAE &
  \cellcolor[HTML]{D9D9D9} &
  \cellcolor[HTML]{CEDD82}3.60 &
  \cellcolor[HTML]{6DC17B}0.03 &
  \cellcolor[HTML]{A4D17E}199 &
  \cellcolor[HTML]{FDD07E}0.30 &
  \cellcolor[HTML]{FEDD81}0.09 &
  \cellcolor[HTML]{FFEB84}0.33 &
  \cellcolor[HTML]{FDC97D}0.49 &
  \cellcolor[HTML]{E1E383}0.34 &
  \cellcolor[HTML]{D9D9D9}N/A &
  \cellcolor[HTML]{D9D9D9}N/A &
  \cellcolor[HTML]{F8696B}156.1 &
  \cellcolor[HTML]{FA8871}131.9 &
  \cellcolor[HTML]{D9D9D9}21.0 \\
Music2Latent (v1) &
   &
  \cellcolor[HTML]{7CC57D}4.01 &
  \cellcolor[HTML]{67BF7B}0.03 &
  \cellcolor[HTML]{B1D47F}238 &
  \cellcolor[HTML]{FDCD7E}0.30 &
  \cellcolor[HTML]{FDD17F}0.08 &
  \cellcolor[HTML]{FEDC81}0.32 &
  \cellcolor[HTML]{FDC77D}0.48 &
  \cellcolor[HTML]{FFEB84}0.27 &
  N/A &
  N/A &
  \cellcolor[HTML]{FFE082}52.9 &
  \cellcolor[HTML]{F8696B}168.7 &
  10.0 \\ %\midrule
\cellcolor[HTML]{D9D9D9}VAE D=64 &
  \cellcolor[HTML]{D9D9D9}recon+KLD+mbGAN &
  \cellcolor[HTML]{FDCA7D}2.76 &
  \cellcolor[HTML]{FFDF82}0.17 &
  \cellcolor[HTML]{FFDD82}582 &
  \cellcolor[HTML]{FEDF81}0.33 &
  \cellcolor[HTML]{FEDC81}0.08 &
  \cellcolor[HTML]{FBB379}0.29 &
  \cellcolor[HTML]{FED880}0.55 &
  \cellcolor[HTML]{FEE883}0.26 &
  \cellcolor[HTML]{D9D9D9}N/A &
  \cellcolor[HTML]{D9D9D9}N/A &
  \cellcolor[HTML]{63BE7B}6.8 &
  \cellcolor[HTML]{63BE7B}4.0 &
  \cellcolor[HTML]{D9D9D9}7.8 \\
VAE D=64 &
  recon+KLD+contr+CLAP+mbGAN &
  \cellcolor[HTML]{FCBE7B}2.55 &
  \cellcolor[HTML]{FED680}0.23 &
  \cellcolor[HTML]{FFEA84}480 &
  \cellcolor[HTML]{ABD380}0.46 &
  \cellcolor[HTML]{CADC81}0.22 &
  \cellcolor[HTML]{E7E483}0.34 &
  \cellcolor[HTML]{7FC77D}0.79 &
  \cellcolor[HTML]{E3E383}0.33 &
  \cellcolor[HTML]{FFEB84}0.08 &
  \cellcolor[HTML]{FFEB84}0.12 &
  \cellcolor[HTML]{63BE7B}6.8 &
  \cellcolor[HTML]{63BE7B}4.0 &
  7.8 \\
\cellcolor[HTML]{D9D9D9}VAE D=128 &
  \cellcolor[HTML]{D9D9D9}recon+KLD+contr+CLAP+mbGAN &
  \cellcolor[HTML]{FCB87A}2.44 &
  \cellcolor[HTML]{FFE784}0.11 &
  \cellcolor[HTML]{FFE383}537 &
  \cellcolor[HTML]{B8D780}0.44 &
  \cellcolor[HTML]{DCE182}0.18 &
  \cellcolor[HTML]{EAE583}0.34 &
  \cellcolor[HTML]{9DCF7F}0.75 &
  \cellcolor[HTML]{FCB87A}0.20 &
  \cellcolor[HTML]{FEDA80}0.07 &
  \cellcolor[HTML]{FFEB84}0.12 &
  \cellcolor[HTML]{63BE7B}6.8 &
  \cellcolor[HTML]{63BE7B}4.0 &
  \cellcolor[HTML]{D9D9D9}7.8 \\
VAE-large D=128 &
  recon+KLD+mbGAN &
  \cellcolor[HTML]{CCDD82}3.61 &
  \cellcolor[HTML]{B8D67F}0.06 &
  \cellcolor[HTML]{F6E883}447 &
  \cellcolor[HTML]{FFEB84}0.36 &
  \cellcolor[HTML]{FFEB84}0.09 &
  \cellcolor[HTML]{FBB178}0.29 &
  \cellcolor[HTML]{FFEB84}0.62 &
  \cellcolor[HTML]{FEDD81}0.25 &
  N/A &
  N/A &
  \cellcolor[HTML]{FFDF82}53.6 &
  \cellcolor[HTML]{FFE784}17.8 &
  7.8 \\
\cellcolor[HTML]{D9D9D9}VAE-large D=128 &
  \cellcolor[HTML]{D9D9D9}recon+KLD+contr+CLAP+mbGAN &
  \cellcolor[HTML]{FFEB84}3.35 &
  \cellcolor[HTML]{FFEB84}0.08 &
  \cellcolor[HTML]{FFEB84}471 &
  \cellcolor[HTML]{8ECB7E}0.49 &
  \cellcolor[HTML]{B4D680}0.27 &
  \cellcolor[HTML]{BFD981}0.37 &
  \cellcolor[HTML]{66BF7C}0.82 &
  \cellcolor[HTML]{C2DA81}0.41 &
  \cellcolor[HTML]{F7E984}0.09 &
  \cellcolor[HTML]{F6E984}0.14 &
  \cellcolor[HTML]{FFDF82}53.6 &
  \cellcolor[HTML]{FFE784}17.8 &
  \cellcolor[HTML]{D9D9D9}7.8 \\ \bottomrule
\end{tabular}%
}
    \vspace{-6pt}
\end{table*}

\subsection{Evaluation tasks and metrics}
We evaluate the VAE along two orthogonal dimensions: \textbf{reconstruction audio quality} and \textbf{latent space representation}.

{ \textbf{For reconstruction quality}: 
     We measure \textit{Speech quality} using DistillMOS~\cite{Stahl2025} on the LibriSpeech test-clean set.
     \textit{Sound quality} is evaluated using the Fréchet Audio Distance (FAD) with \ac{CLAP} embeddings~\cite{Elizalde2024}, computed on permissively licensed samples from MUSDB18.
     \textit{Speech content preservation} is quantified using Word Error Rate (WER) with Whisper Large v3.

\textbf{For latent space representation}:
We probe the latent space by training simple MLP classifiers on the learned representations for several downstream tasks:
     \textit{Audio scene classification} (TAU Urban Acoustic Scenes),
    \textit{Multi-label sound event detection} (FSD50k~\cite{Fonseca2021}),
     \textit{Speech emotion recognition} (MSP-Podcast v1.10~\cite{Lotfian2019}),
     \textit{Music genre classification} (GTZAN~\cite{Tzanetakis2002}),
     \textit{Musical instrument detection} (NSynth~\cite{Engel2017}).
All classification results are reported using mean Average Precision (mAP).\\
We additionally evaluate for \textbf{zero-shot classification} capabilities for models trained with a CLAP loss, using the same classification test sets as for latent space representation. Finally, we assess \textbf{audio captioning} on datasets AudioCaps and Clotho using metric SPIDEr.
% The VAE is evaluated along two orthogonal dimensions, \textbf{reconstruction audio quality} and \textbf{latent space representation}.
% To evaluate the reconstruction quality, for speech quality we use DistillMOS \cite{Stahl2025} on Librispeech test-clean, for sound quality \ac{FAD} with \ac{CLAP} \cite{Elizalde2024} embeddings on permissively licensed samples from Musdb18. Speech content preservation is assessed by \ac{WER} using Whisper Large v3.

% The latent space is probed by training simple MLP classifiers on these features for several tasks. We use audio scene classification (TAU-Urban), multi-label sound event detection (FSD50k \cite{Fonseca2021}), speech emotion recognition (MSP-Podcast \cite{Lotfian2019} v1.10) and music genre (GTZAN \cite{Tzanetakis2002}) and music instrument detection (Nsynth \cite{Engel2017}), all reported in \ac{mAP}.
% % IEMOCAP \cite{Busso2008}
% We further use also evaluate the ability of zero-shot classification for models including CLAP loss on the same classification test sets.
% The audio captioning ability is evaluated on AudioCaps and Clotho using SPIDEr.

\subsection{Baselines}
As baselines we use existing continuous latent space autoencoders from StableAudio \cite{Evans2025}, Music2Latent \cite{Pasini2024}.

For latent space evaluation tasks, we also add the CLAP audio encoder \cite{Elizalde2024} as reference, which however cannot generate sound.
Note that in the Music2Latent paper \cite{Pasini2024}, the authors evaluated their latent space \emph{before} the bottleneck, i.\,e., using a much larger feature space, which improves performance, but makes direct comparison with other VAEs difficult, as it evaluates a higher-dimensional representation than the actual bottleneck latent space. In this study, we train the classifiers on the actual low-dimensional latent space in the bottleneck for all autoencoders. Further, Music2Latent uses a transformer architecture and therefore does not scale to arbitrary sequence lengths. The published model operates on 1~s chunks for efficiency and creates sometimes notable stitching artifacts.

\subsection{Results}
Table~\ref{tab:loss_ablation} shows the contribution of each loss component to the information density of the latent space. The first row shows the base VAE model only with reconstruction loss and \ac{KLD}. The next 3 rows show that the contrastive loss and CLAP loss improve classification results, most significantly the CLAP loss. Combining CLAP and contrastive loss yields another improvement and is the strongest model on latent space probing. The reconstruction quality is low without the adversarial loss (first 4 VAE models), yielding DistillMOS below 2, high WER and high FAD. The most audible effect is failure to reconstruct high frequencies. Further, it is notable that that adding the adversarial loss improves not only significantly the reconstruction quality, but also latent space representation compared to the base VAE with only recon+KLD loss. The VAE with adversarial, but without semantic losses yields the highest reconstruction quality metrics. 
We trained a model with recon+KLD+mbGAN, but without the training scheme to enhance audio (i.\,e.\, no denoising autoencoder principle) as described in Sec.~\ref{sec:data_augmentation}, where we replace the target signal $\hat{\mathbf{x}}$ with the degraded signal $\hat{\mathbf{y}}$ in \eqref{eq:denoising_vae} for training. We can see that without enhancement, all reconstruction metrics drop, which demonstrates its effectiveness.

Interestingly, combining all losses results in a minor degradation in both reconstruction quality (compared to best model with adversarial but no semantic losses)
and latent space representation (best model with semantic losses but no adversarial loss). However, combining all losses balances both properties and still maintains strong performance across all metrics.

Table~\ref{tab:loss_ablation} also shows zero-shot classification ability for models trained with the CLAP loss. As expected, zero-shot classification does not reach the performance of the supervised trained MLPs, but still achieves competitive results across the four classification tasks, indicating strong generalization. This new ability opens promising avenues for applications of the proposed VAE compared to existing methods without zero-shot audio-text capabilities. 

Table~\ref{tab:results_overall} summarizes overall performance compared to baselines. We present our VAE in 3 different model architecture configurations, a small model with latent size $D\!=\!64$, a small model (same parameter count) with increased latent size $D\!=\!128$, and a large model (increased parameter count) with latent size $D\!=\!128$.
The upper bound for reconstruction quality is the original audio, while CLAP serves as an upper reference for captioning, since the VAEs distill the CLAP embeddings. Note that CLAP is not able to generate audio, so direct comparison to the VAEs is not intended.
While StableAudio VAE and Music2Latent are strong baselines for reconstruction, StableAudio is over 10$\times$ larger and more complex, and Music2Latent is similarly over 10$\times$ larger than our small model - similar parameter count but yet still 10$\times$ more complex than our largest model. Notably, our VAEs operate at the lowest latent frame rate among all compared models.
The small VAE models with $D\!=\!64$ achieve acceptable audio fidelity, but do not reach the strong baselines. However, the VAE $D\!=\!64$ with all losses outperforms all baseline codecs in terms of latent space probing, and is able to caption. Enlarging the latent dimension to 128 improved WER, but not DistillMOS and FAD, and no significant change in latent space strength. 
Only scaling up the VAE architecture significantly boosts the audio fidelity, reaching comparable performance to StableAudio and Music2Latent, while outperforming them on all latent space tasks. The SALAD-VAE configuration in the last row performs well across the board on all metrics. Also for the large VAE model, removing the semantic losses (contrastive and CLAP) mildly boosts the reconstruction fidelity further, at the cost of latent space performance and losing captioning ability. Finally, unlike CLAP which operates on fixed 7~s segments, our model supports arbitrary-length audio and produces time-variant embeddings, enabling more flexible downstream applications.

\section{Conclusion}
We proposed a general purpose audio VAE that achieves strong performance across diverse audio types -- speech, music, and general sounds -- all while maintaining high reconstruction fidelity and a compact, information-dense latent space. The architecture is practical for processing arbitrary-length audio and has significantly lower complexity than comparable models. We showed that latent space information density improves with the proposed contrastive and CLAP losses. Moreover, the distillation of text-audio embeddings enables caption generation and zero-shot classification via the CLAP text decoder capabilities, a property that has not been previously demonstrated in audio codecs. Future work includes extending the model to multi-channel audio formats.

%\begin{table}[htbp]
%\caption{Table Type Styles}
%\begin{center}
%\begin{tabular}{|c|c|c|c|}
%\hline
%\textbf{Table}&\multicolumn{3}{|c|}{\textbf{Table Column Head}} \\
%\cline{2-4} 
%\textbf{Head} & \textbf{\textit{Table column subhead}}& \textbf{\textit{Subhead}}& \textbf{\textit{Subhead}} \\
%\hline
%copy& More table copy$^{\mathrm{a}}$& &  \\
%\hline
%\multicolumn{4}{l}{$^{\mathrm{a}}$Sample of a Table footnote.}
%\end{tabular}
%\label{tab1}
%\end{center}
%\end{table}

%\begin{figure}[htbp]
%\centerline{\includegraphics{fig1.png}}
%\caption{Example of a figure caption.}
%\label{fig}
%\end{figure}

%\section*{Acknowledgment}
%
%The preferred spelling of the word ``acknowledgment'' in America is without 
%an ``e'' after the ``g''. Avoid the stilted expression ``one of us (R. B. 
%G.) thanks $\ldots$''. Instead, try ``R. B. G. thanks$\ldots$''. Put sponsor 
%acknowledgments in the unnumbered footnote on the first page.

\newpage
%\balance

%\section*{References}
\bibliographystyle{IEEEbib-abbrev}
\bibliography{./sapref.bib}

%column names are read directly from the csv. To change how column names appear, it's a 2-step process: 1)change them in the .csv column names, 2) replace the correct names in file table1_loss_ablation.tex, in places where "getcolminmax" is found, and where "columns" is found
% \include{utils/table1_loss_ablation.tex}

\end{document}